\documentclass[%
reprint,
superscriptaddress,
amsmath,amssymb,
aps,
prstab,
]{revtex4-2}
\usepackage{xcolor}
\usepackage{graphicx}
\usepackage{dcolumn}
\usepackage{bm}
\usepackage{booktabs}
\usepackage{multirow}
\usepackage{amssymb}
\usepackage{float}
\usepackage{graphicx}
\usepackage{epstopdf}
\usepackage{epsfig} 

\begin{document}

\title{ Magnetic-field oscillations of the critical temperature in ultraclean, two-dimensional Type-I superconductors}
\author{Aiying Zhao}
\email{ayzhao0909@sina.cn}
\affiliation{Institute of Theoretical Physics, University of Science and Technology Beijing, Beijing 100083, People’s Republic of China}
\author{Richard A Klemm}
\email{richard.klemm@ucf.edu}
\affiliation{Department of Physics, University of Central Florida, Orlando, FL 32816-2385, United States of America}
\author{Qiang Gu}
\email{qgu@ustb.edu.cn}
\affiliation{Institute of Theoretical Physics, University of Science and Technology Beijing,  Beijing 100083, People’s Republic of China}

\date{\today}

\baselineskip12pt

\begin{abstract}
We investigate the influence of Landau Levels (LLs) and Zeeman energy, induced by an applied magnetic field ${\bf B}$, on the critical temperature $T_c$ for two-dimensional (2D) ultraclean metals using a fully quantum mechanical approach within the Bardeen-Cooper-Schrieffer (BCS) theory. In contrast to standard BCS theory, it allows for Cooper pair formation between electrons with opposite spins and momenta along the ${\bf B}$ direction, both on the same or on neighboring LLs. Our quantum mechanical treatment of LLs reveals that $T_c({\bf B})$ for electrons paired on the same LLs exhibits oscillations around the BCS critical temperature at lower magnetic fields, a phenomenon analogous to the de Haas-van Alphen effect. The Zeeman energy leads to a decrease in $T_c({\bf B})$ with increasing ${\bf B}$ for electrons paired both on the same and on neighboring LLs. Notably, as the $g$-factor increases, the amplitude of the ${\bf B}$ oscillations gradually diminishes until they vanish at higher magnetic fields. Conversely, for small $g$-factors, electron pairing on the same or on neighboring LLs can result in a re-entrant superconducting phase at very high magnetic fields.
\end{abstract}

 \vskip0pt

\maketitle

\section{Introduction}
A paramount characteristic of superconductors is their critical temperature ($T_c$). Understanding the fundamental physics of superconductivity relies heavily on both experimental measurements and theoretical investigations of this key property. Ongoing research continues to explore the diverse factors that influence the critical temperature, including isotope effects \cite{BCS-1957,Gweon-2004}, impurities \cite{Hirschfeld-2009,Markowitz-1963}, pressure \cite{Somayazulu-2019,Mozaffari-2019,Drozdov-2019}, and magnetic fields \cite{Drozdov-2019,Rasolt-1989,Rasolt-1991,Gruenberg-1968,Rajagopal-1966,Gruenberg-1966,Rasolt-1992,Norman-1992,MacDonald-1992,Chaudhary-2021,MacDonald-1993}. Among these influential factors, the impact of magnetic fields on the superconducting critical temperature remains a crucial and actively investigated area of research, particularly concerning phenomena such as the upper critical fields \cite{Ran-2019} and vortex states \cite{Abrikosov-2004}. In this study, we specifically focus on how the magnetic field affects the critical temperature in two-dimensional (2D) Type I superconductors.

To understand the behavior of electrons in a magnetic field, researchers often employ the semiclassical approximation, especially for the upper critical field near $T_c$ \cite{WHH-1964,WHH2-1996,WHH3-1996,Zhao-2022}. This approach is effective when the magnetic field varies gradually compared to the electron's wavelength. Near $T_c$, this is further supported by the Ginzburg-Landau (GL) theory. In this approximation, Cooper pairs are treated using normal state properties, and the magnetic field's effect on the electrons is mainly captured by a phase factor, simplifying the calculations. Under certain conditions, this semiclassical approximation can reveal behavior consistent with Cooper pairs occupying the lowest Landau levels (LLs). 

However, this semiclassical picture breaks down in the presence of a strong magnetic field {\bf B} and near the superconducting-to-normal state transition, where the assumption of treating electrons forming Cooper pairs as simple normal-state electrons (without considering Landau quantization) becomes increasingly inadequate. Consequently, many studies have delved into the influence of LLs on superconductors at very high magnetic fields, but often focused on the intriguing “quantum” limit where only the lowest LL or a few LLs are occupied \cite{Rasolt-1991,Rasolt-1992}, a regime known to exhibit particularly interesting physical phenomena such as the reentrance of superconductivity. However, achieving this quantum limit experimentally remains a significant challenge for most conventional superconductors, typically requiring magnetic fields on the order of thousands of teslas, which are often difficult to attain. While recent research indicates that magic-angle twisted bilayer graphene (MATBG), a low-carrier-density system, offers the potential to reach this quantum limit at more accessible field strengths \cite{Chaudhary-2021}, the significantly higher carrier density found in most superconductors means that they typically do not reach this extreme quantum limit under experimentally feasible conditions. Instead, these materials exhibit thousands of LLs near to the Fermi surface (FS), making the investigation of LL effects in this multi-level regime a more common area of research.

Beyond the orbital quantization effects captured by LLs, Zeeman splitting (Zeeman energy) also plays a crucial role in the behavior of superconductors in a magnetic field, acting as another significant factor that suppresses superconductivity. A significant consequence of the Zeeman splitting is the determination of the Pauli limit ($B_p$) \cite{Chandrasekhar-1962,Clogston-1962}, which represents the magnetic field strength above which spin polarization energy overcomes the superconducting condensation energy. However, here, we have chosen to neglect the effect of the Pauli limit, which implies that the high magnetic fields considered will not lead to significant spin polarization of the electrons. Indeed, while Coulomb interactions, spin-spin interactions, and magnetic interactions in a magnetic field can all affect the electron behavior \cite{Tu-2018, Yang-2024}, we have opted to neglect these complexities to maintain a simplified yet insightful physical model.

Building upon the understanding of the magnetic field effects, we know that electrons on the same LL with opposite spin orientations possess distinct energies due to Zeeman splitting. The energy separation between LLs in a magnetic field, combined with even a small attractive interaction, can induce superconductivity in 2D materials. Assuming spin singlet pairing, this opens up the possibility of various Cooper pairing configurations, including pairing between electrons on the same LL, between adjacent LLs, or across mixed LLs. To simplify our investigation, we focused on the formation of Cooper pairs by electrons with opposite spin occupying either the same or neighboring LLs with opposite spin (as illustrated in Figure 1). 

To analyze this specific scenario, we employed a fully quantum mechanical approach within the framework of BCS theory, in contrast to the semiclassical approach. The remainder of this paper is structured as follows: Section II introduces the quantized Hamiltonian for electrons paired on both the same and neighboring LLs, and subsequently derives the critical temperatures for these two models. Section III then presents and discusses the numerical results obtained from the equations in Section II, detailing the effects of the LLs and the Zeeman splitting. Finally, Section IV summarizes the main findings and conclusions of this study.

\section{Model}
\subsection{Electrons paired on the same LL}

\begin{figure}
\centering
{\includegraphics[width=0.45\textwidth]{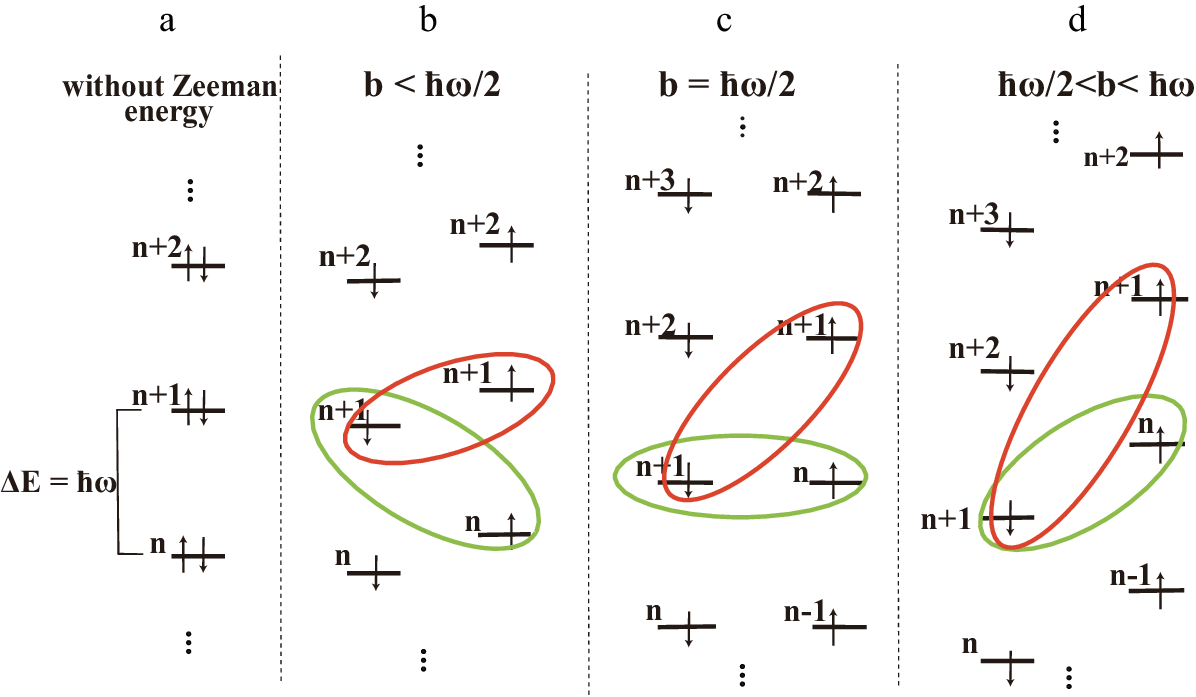}
\caption{ A schematic diagram of LLs and electron pairing in 2D. The arrows depict the electron spins, $n$, $n+1$... represent the number of LLs in a thin shell around the FS, 
$b$ is the Zeeman energy, $\hbar\omega$ is the energy difference between the nearest neighboring LLs. The first column (a) doesn't include the Zeeman energy, and the last three columns (b-d) show three different cases of the Zeeman energy relative to one half of the energy difference between the neighboring LLs. The red circles in (b-d) illustrate electrons paired on the same LLs, whereas the green circles represent electrons paired on neighboring LLs. In the figures, the energy of the electrons with up spins in the $n$th LL is $n\hbar\omega+b$, and the energy of electrons with down spins electrons in the $(n+1)$th LL is $(n+1)\hbar\omega-b$. Figure c specifically depicts the case where electrons paired on neighboring LLs have equal energy.}
\label{fig1}}
\end{figure}

The Hamiltonian for electrons in a 2D superconductor in a perpendicular magnetic field B on an ellipsoidal Fermi surface and paired on the same LL is
\begin{eqnarray}
H&=&\sum _{n}\frac{eB L^{2} }{2\pi \hbar} 
 \biggl( (\epsilon_{n}+b ) a^{\dagger}_{n,\uparrow}a_{n\uparrow} +(\epsilon_{n}-b ) a^{\dagger}_{n,\downarrow}a_{n,\downarrow}
    \biggr) \nonumber\\
&& -V_{int}\sum_{n}  \frac{eB L^{2}}{2\pi \hbar}  (a^{\dagger}_{n,\uparrow}a^{\dagger}_{n,\downarrow}  a_{n,\downarrow}a_{n,\uparrow}),
\end{eqnarray}
\begin{eqnarray}
\epsilon_{n} = (n+\frac{1}{2})\hbar\omega-\mu_F,
\end{eqnarray}
in which $-V_{int}$ is a small attractive electron-electron interaction for electronic energies within a shell around the FS,  $\hbar$=$\frac{h}{2\pi}$ is the reduced Planck constant,  $\omega=\frac{eB}{m_{xy}}$ is the cyclotron frequency, $m_{xy}=\sqrt{m_x^2+m_y^2}$ represents the effective masses in the plan \cite{KC,KC-1993,KC-1994,KC-2012},  $b=|\frac{g}{2} \mu_B \bm{\sigma}\cdot \bm{B}| $ is the Zeeman energy, $\frac{eBL^2}{2\pi\hbar}$ is the degeneracy of a LL, and $L^2$ is the area of the sample normal to ${\bf B}$. The summation index $n$ is restricted by the condition $ -\hbar\omega_D\leq(n+\frac{1}{2})\hbar\omega\pm b-\mu_F\leq \hbar\omega_D$.

To diagonalize the Hamiltonian, we employ a modified Bogoliubov-Valatin transformation \cite{Bogoljubov-1958,Valatin-1958} by defining the new fermionic operators as
\begin{align}
  \gamma_{n,\uparrow}&=\mu_{n}a_{n,\uparrow}-\nu_{n}a^{\dagger}_{n,\downarrow},\nonumber\\
  \gamma_{n,\downarrow}&=\mu_{n}a_{n,\downarrow}+\nu_{n}a^{\dagger}_{n,\uparrow}; \nonumber\\
  \gamma^{\dagger}_{n,\uparrow}&=\mu^{\ast}_{n}a^{\dagger}_{n,\uparrow}-\nu^{\ast}_{n}a_{ n,\downarrow},\nonumber\\
  \gamma^{\dagger}_{n,\downarrow}&=\mu^{\ast}_{n}a^{\dagger}_{n,\downarrow}+\nu^{\ast}_{n}a_{n,\uparrow},
\end{align}
where the coefficients $\mu_{n}$, $\nu_{n}$ satisfy
\begin{align}
  |\mu_{n}|^2+|\nu_{n}|^2=1.
\end{align}

Then it is easy to obtain
\begin{align}
  a_{n,\uparrow}&=\mu^{\ast}_{n} \gamma_{n,\uparrow} + \nu_{n}\gamma^{\dagger}_{n,\downarrow},\nonumber\\
  a_{n,\downarrow}&=\mu^{\ast}_{n}\gamma_{n,\downarrow}-\nu_{n}\gamma^{\dagger}_{n,\uparrow}; \nonumber\\
  a^{\dagger}_{n,\uparrow}&=\mu_{n} \gamma^{\dagger}_{n,\uparrow} + \nu^{\ast}_{n}\gamma_{n,\downarrow},\nonumber\\
  a^{\dagger}_{n,\downarrow}&=\mu_{n}\gamma^{\dagger}_{n,\downarrow}-\nu^{\ast}_{n}\gamma_{n,\uparrow}.
\end{align}

After substituting equation (5) into equation (1), the Hamiltonian becomes
\begin{align}
 H=H_0+H_1+E_g,
\end{align}
where
\begin{align}
H_0&= \frac{eBL^2 }{2\pi \hbar}  \sum _{n}
 \Biggl[ \biggl(  (\epsilon_{n}+b )|\mu_{n}|^2 -(\epsilon_{n}-b)|\nu_{n}|^2\nonumber\\
&  +\mu^{\ast}_{n} \nu_{n}\Delta^{\ast} + \mu_{n} \nu^{\ast}_{n} \Delta    \biggr)
  \gamma^{\dagger}_{n,\uparrow} \gamma_{n,\uparrow}  \nonumber\\
  & +\biggl(  -(\epsilon_{n}+b )|\nu_{n}|^2 +(\epsilon_{n}-b)|\mu_{n}|^2
  + \mu^{\ast}_{n} \nu_{n}\Delta^{\ast} \nonumber\\
  &+ \mu_{n} \nu^{\ast}_{n} \Delta    \biggr)
  \gamma^{\dagger}_{n,\downarrow} \gamma_{n,\downarrow}  \Biggr],  \nonumber\\
 H_1&=\frac{eBL^2}{2\pi \hbar}  \sum_{n}
 \Biggl[ \biggl(  2\epsilon_{n} \mu_{n}\nu_{n}+ \Delta^{\ast}\nu_{n}^2
 -\Delta \mu_{n}^{2}   \biggr)
  \gamma^{\dagger}_{n,\uparrow}\gamma^{\dagger}_{n,\downarrow}  \nonumber\\
 & -\biggl(  2\epsilon_{n} \mu^{\ast}_{n}\nu^{\ast}_{n}+ \Delta \nu_{n}^{\ast 2}
  -\Delta^{\ast} \mu^{\ast 2}_{n}   \biggr)
   \gamma_{n,\uparrow} \gamma_{n,\downarrow}   \Biggr], \nonumber\\
E_g &=\frac{eBL^2 }{2\pi \hbar}  \sum_{n}
  \biggl(   2\epsilon_{n} |\nu_{n}|^2  -\Delta^{\ast} \mu^{\ast}_{n} \nu_{n} 
   -\Delta \mu_{n} \nu^{\ast}_{n}   \biggr)
  +\frac{|\Delta|^2}{V_{int}},
\end{align}
where the superconducting energy gap $\Delta(T)$  satisfies
\begin{align}\label{Gap}
  \Delta(T)=V_{int}\sum_{n} \frac{eBL^2 }{2\pi\hbar}  < a_{n,\downarrow} a_{n,\uparrow} >.
\end{align}

Since $H_1$ just includes off-diagonal terms in the $\gamma$ and $\gamma^{\dag}$ operators, we require it  to  vanish, leading to
\begin{align}
2\epsilon_{n} \mu_{n}\nu_{n}+ \Delta^{\ast}\nu_{n}^2-\Delta \mu_{n}^{2} =0,\nonumber\\
2\epsilon_{n} \mu^{*}_{n}\nu^{*}_{n}+ \Delta\nu_{n}^{*2 }-\Delta^{*} \mu_{n}^{*2} =0.
\end{align}

From  equation (4) and the two lines of equation (9), we obtain:
\begin{align}
  &|\mu_{n}|^2=\frac{1}{2}(1+\frac{\epsilon_{n}}{\xi_{n}}),\nonumber\\
  &|\nu_{n}|^2=\frac{1}{2}(1-\frac{\epsilon_{n}}{\xi_{n}}),
\end{align}
where $\xi_{n}=\sqrt{|\Delta|^2 + \epsilon^{2}_{n} }$.
Now the Hamiltonian, equation (1),  is diagonalized
\begin{align}
 H&=H_0+E_g \nonumber\\
  &=\frac{eBL^2}{2\pi \hbar}  \sum_{n}
 \Biggl[ ( \xi_{n}+b )  \gamma^{\dagger}_{n,\uparrow} \gamma_{n,\uparrow} 
        +( \xi_{n}-b ) \gamma^{\dagger}_{n,\downarrow} \gamma_{n,\downarrow}  \Biggr] \nonumber\\
        &+\frac{eBL^2 }{2\pi \hbar}  \sum_{n}  (\epsilon_{n}-\xi_{n})+\frac{\Delta^2}{V_{int}}.
\end{align}

Using the newly defined fermionic operators in equation (5), we may rewrite the above self-consistent  equation (8) as
\begin{align}
1=V_{int}\frac{eBL^2 }{2\pi\hbar} \sum _{n}
         \frac{  1- \frac{1}{1+ e^{\frac{ \xi_{n} +b  }{k_{B} T}}} -\frac{1}{1+ e^{\frac{ \xi_{n} -b  }{k_{B} T}}} }{2\xi_{n}}.  
\end{align}

When $\Delta(T)\rightarrow 0$, then $\xi_{n}\rightarrow |\epsilon_{n}|$, the elementary excitation of the superconducting state reduces to that of the normal state. We obtain the critical temperature $T_c(B)$ in equation (12) by setting $\Delta(T)=0$, yielding
\begin{align}
1=V_{int}\frac{eBL^2 }{2\pi \hbar}  \sum_{n} &\biggl[ \frac{ \sinh  \frac{  |( n+\frac{1}{2})\hbar\omega -\mu_F |}{k_B T_c}   }
{\cosh \frac{b}{k_B T_c}  + \cosh\frac{   |( n+\frac{1}{2})\hbar\omega -\mu_F|}{k_B T_c}  } \nonumber\\
&\times\frac{1}{2|( n+\frac{1}{2})\hbar\omega -\mu_F|}\biggr].
\end{align}

\subsection{Electrons paired on neighboring LLs}

The Hamiltonian of electrons paired on neighboring LLs is
\begin{align}
H^{\prime}=&\sum_{n}\frac{eBL^2}{2\pi\hbar}  \biggl[ (\epsilon_{n}+b)  a^{\dagger}_{n,\uparrow}   a_{n,\uparrow}
   + (\epsilon_{n+1}-b) a^{\dagger}_{n+1,\downarrow} a_{n+1,\downarrow}   \biggr]  \nonumber\\
 &-\sum_{n}\frac{eBL^2 }{2\pi\hbar}  \biggl(   \Delta^{\ast} a_{n,\uparrow} a_{n+1,\downarrow} 
  + \Delta   a^{\dagger}_{n+1,\downarrow}a^{\dagger}_{n,\uparrow}           \biggr)
 +\frac{|\Delta|^2}{V_{int}},
\end{align}
the index $n$ in the summation is restricted by the condition $ -\hbar\omega_D\leq(n+\frac{1}{2})\hbar\omega+ b-\mu_F\leq \hbar\omega_D$,
$ -\hbar\omega_D\leq(n+\frac{3}{2})\hbar\omega- b-\mu_F\leq \hbar\omega_D$, and $\Delta(T)$ satisfies
\begin{align}
   \Delta(T)=V_{int}\sum_{n} \frac{eBL^2 }{2\pi\hbar}  < a_{n,\uparrow} a_{n+1,\downarrow} >
\end{align}

We then employ a modified Bogoliubov transformation by defining the new fermionic operators:
\begin{align}
  \gamma_{n,\uparrow}&=\overline{\mu}_{n}a_{n,\uparrow}+\overline{\nu}_{n}a^{\dagger}_{n+1,\downarrow},\nonumber\\
  \gamma_{n+1,\downarrow}&=\overline{\mu}_{n}a_{n+1,\downarrow}-\overline{\nu}_{n}a^{\dagger}_{n,\uparrow}; \nonumber\\
  \gamma^{\dagger}_{n,\uparrow}&=\overline{\mu}_{n}a^{\dagger}_{n,\uparrow}+\overline{\nu}_{n}a_{n+1,\downarrow},\nonumber\\
  \gamma^{\dagger}_{n+1,\downarrow}&=\overline{\mu}_{n}a^{\dagger}_{n+1,\downarrow}-\overline{\nu}_{n}a_{n,\uparrow},
\end{align}
where the coefficients $\overline{\mu}_{n}$, $\overline{\nu}_{n}$ are real and satisfy
\begin{align}
  \overline{\mu}_{n}^2+\overline{\nu}_{n}^2=1.
\end{align}
Then it is easy to obtain
\begin{align}
  a_{n,\uparrow} &= \overline{\mu}_{n} \gamma_{n,\uparrow} - \overline{\nu}_{n}  \gamma^{\dagger}_{n+1,\downarrow},\nonumber\\
  a_{n+1,\downarrow}&=\overline{\mu}_{n}\gamma_{n+1,\downarrow} + \overline{\nu}_{n} \gamma^{\dagger}_{n,\uparrow}; \nonumber\\
  a^{\dagger}_{n,\uparrow}&=\overline{\mu}_{n} \gamma^{\dagger}_{n,\uparrow} - \overline{\nu}_{n}\gamma_{n+1,\downarrow},\nonumber\\
  a^{\dagger}_{n+1,\downarrow}&=\overline{\mu}_{n}\gamma^{\dagger}_{n+1,\downarrow}+\overline{\nu}_{n}\gamma_{n,\uparrow}.
\end{align}
After substituting equation (18) into equation (14), the effective Hamiltonian becomes
\begin{align}
H^{\prime}=H_{0}^{\prime}+H_{1}^{\prime}+E^{\prime}_g ,
\end{align}
where
\begin{align}
H_{0}^{\prime}=& \frac{eBL^2 }{2\pi \hbar}  \sum_{n}
 \Biggl[ \biggl(  (\epsilon_{n}+b )\overline{\mu}_{n}^2  
 -(\epsilon_{n+1}-b)\overline{\nu}_{n}^2 \nonumber\\
& + 2\overline{\mu}_{n} \overline{\nu}_{n}\Delta     \biggr)
  \gamma^{\dagger}_{n,\uparrow} \gamma_{n,\uparrow}  
   +\biggl(  -(\epsilon_{n}+b )\overline{\nu}_{n}^2 \nonumber\\
&+(\epsilon_{n+1}-b)\overline{\mu}_{n}^2
  + 2\overline{\mu}_{n} \overline{\nu}_{n}\Delta     \biggr)
  \gamma^{\dagger}_{n+1,\downarrow} \gamma_{n+1,\downarrow}  \Biggr],  \nonumber\\
H_{1}^{\prime}=&\frac{eBL^2 }{2\pi \hbar}  \sum_{n}
 \Biggl( (\epsilon_{n}+\epsilon_{n+1}) \overline{\mu}_{n}\overline{\nu}_{n} 
 + \Delta(\overline{\nu}_{n}^2- \overline{\mu}_{n}^{2})    \Biggr) \nonumber\\
& \times
  (\gamma^{\dagger}_{n+1,\downarrow}\gamma^{\dagger}_{n,\uparrow} 
 +   \gamma_{n,\uparrow} \gamma_{n+1,\downarrow}  ), \nonumber\\
E_{g}^{\prime}=&\frac{eBL^2 }{2\pi \hbar}  \sum_{n}
   \biggl(   (\epsilon_{n} + \epsilon_{n+1} )\overline{\nu}_{n}^2 
   -2\Delta \overline{\mu}_{n} \overline{\nu}_{n}   \biggr)
  +\frac{|\Delta|^2}{V_{int}}.
\end{align}
Since $H_{1}^{\prime}$ includes off-diagonal terms in the $\gamma$ and $\gamma^{\dag}$ operators,  we require that it vanishes, implying that
\begin{align}
 (\epsilon_{n}+\epsilon_{n+1}) \overline{\mu}_{n}\overline{\nu}_{n} + \Delta(\overline{\nu}_{n}^2- \overline{\mu}_{n}^{2})  =0.
\end{align}
Combining  equations (17) and (21), we obtain:
\begin{align}
  &\overline{\mu}_{n}^2=\frac{1}{2}\Biggl(1+\frac{\epsilon_{n,n+1} }{\xi_{n,n+1}  } \Biggr),\nonumber\\
  &\overline{\nu}_{n}^2=\frac{1}{2}\Biggl(1-\frac{\epsilon_{n,n+1} }{\xi_{n,n+1}  } \Biggr),
\end{align}
where $\xi_{n,n+1}=  \sqrt{ \Delta^2 +\epsilon_{n,n+1}^2}$, and 
\begin{align}
\epsilon_{n,n+1}=\frac {\epsilon_{n} + \epsilon_{n+1}}{2}.
\end{align}
Note that due to equation (21), 
$\overline{\mu}_{n}$ and $\overline{\nu}_{n}$ are functions of $\epsilon_{n,n+1}$ and $\Delta$.
Finally, the Hamiltonian for electrons paired on neighboring LLs is diagonalized and becomes 
\begin{align}
 H^{\prime}&=H_{0}^{\prime}+E_{g}^{\prime}\nonumber \\
  &=\frac{eBL^2 }{2\pi \hbar}  \sum_{n}
 \Biggl[ \biggl(  \xi_{n,n+1} -\epsilon_{n,n+1} + b   \biggr)  \gamma^{\dagger}_{n,\uparrow} \gamma_{n,\uparrow}\nonumber\\
   &+\biggl(  \xi_{n,n+1}  +\epsilon_{n,n+1} - b    \biggr)
  \gamma^{\dagger}_{n+1,\downarrow} \gamma_{n+1,\downarrow}  \Biggr]  \nonumber \\
  & +\frac{eBL^2 }{2\pi \hbar}  \sum_{n}  (\epsilon_{n,n+1} -\xi_{n,n+1} )+\frac{\Delta^2}{V_{int}}.
\end{align}

Using the newly defined fermionic operators in equation (18), we may rewrite the self-consistent equation (15) as
\begin{align}
1=V_{int}\frac{eBL^2 }{2\pi\hbar} \sum_{n}
\frac{ 1- \frac{1}{1+ e^{\frac{ \xi_{n,n+1} +b  }{k_{B} T}}} -\frac{1}{1+ e^{\frac{ \xi_{n,n+1} -b  }{k_{B} T}}} }
{ 2\xi_{n,n+1}} 
\end{align}

When $\Delta(T)\rightarrow 0$, then $\xi_{n,n+1}\rightarrow |\epsilon_{n,n+1}|$, the elementary excitations of the superconducting state reduce to those of the normal state. We obtain the critical temperature $T_c(B)$ from  equation (25) by setting $\Delta(T)=0$, 
\begin{align}
1=V_{int}\frac{eBL^2 }{2\pi \hbar}  \sum_{n} & \frac{ \sinh  \frac{  |( n+1)\hbar\omega -\mu_F }{k_B T_c}|   }
{\cosh \frac{b}{k_B T_c}  + \cosh\frac{  |( n+1)\hbar\omega -\mu_F|}{k_B T_c}  } \nonumber\\
&\times \frac{1}{2|( n+1)\hbar\omega -\mu_F|}.
\end{align}

\section{Numerical results and Discussion}

This work investigates the influence of LLs and Zeeman splitting on the superconducting critical temperature ($T_c$) within the framework of BCS theory. To establish a baseline scenario, we consider a 2D model with Fermi energy $\mu_F=0.1$ eV and a Debye frequency of $\hbar\omega_D=0.01$ eV, which yields a BCS critical temperature ($T_c$) of 1 K. The BCS attractive interaction strength, $-V_{int}$, is determined by the relation $k_{B} T_{c}=(2e^{\gamma}/\pi)\hbar\omega_D \exp(-1/N(0)V_{int})$, where $\gamma \approx 0.5772$ is Euler's constant and $2e^{\gamma}/\pi \approx 1.13$. To isolate the effects of other parameters on $T_c$, we maintain the $-V_{int}$ constant across all simulations. Here, $N(0)$ represents the electronic density of states at $\mu_F$. Initially, we present numerical results for the critical temperature in magnetic fields for a fully isotropic system, characterized by $b = \hbar\omega/2$ and $g = 2$. We employed the Bogoliubov-Valatin transformation to diagonalize the Hamiltonian (equations (1) and (14)) and to determine the critical temperature in magnetic fields $T_c({\bf B})$ by assuming that the gap vanishes in the self-consistent equations (8) and (15) at the critical temperature $T_c$ under a magnetic field. 

\begin{figure}
\centering
{\includegraphics[width=0.5\textwidth]{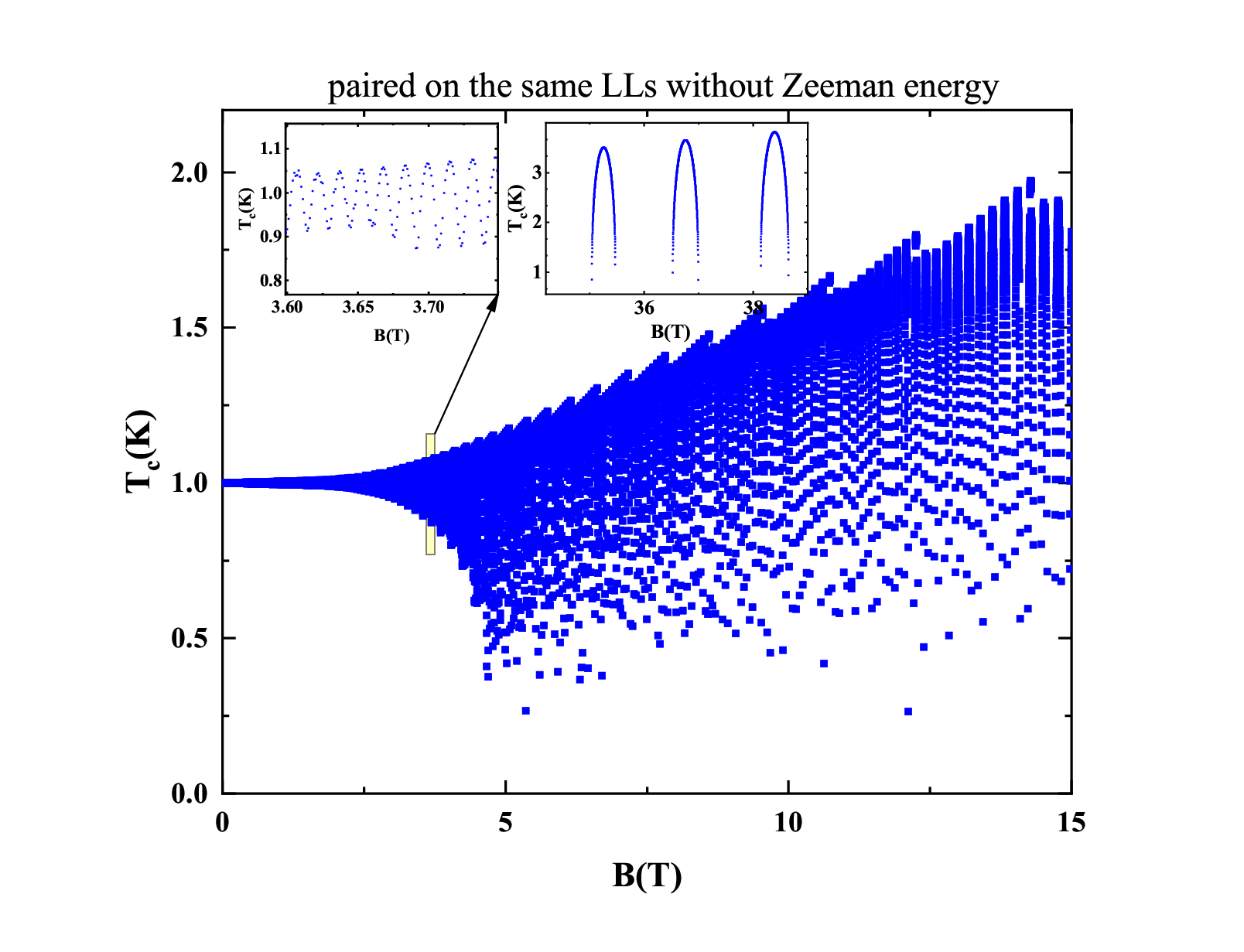}
\caption{ The critical temperature $T_c({\bf B})$ exhibits an oscillatory behavior in a magnetic field for a fully isotropic system, where electrons pair on the same LL without Zeeman splitting. The inset clearly shows the LL-included oscillations of $T_c({\bf B})$ when Zeeman splitting is neglected. }
\label{fig2}}
\end{figure}

To first understand the role of LLs, we considered the superconducting critical temperature in a magnetic field for a fully isotropic system, ignoring Zeeman splitting. Figure 2 illustrates the magnetic field dependence of this critical temperature, $T_c({\bf B})$, specifically for electron pairing on the same LLs with opposite spins. At low magnetic fields, $T_c({\bf B})$ is approximately 1 K, consistent with standard BCS theory. In this regime, numerous LLs are populated near the Fermi surface, and the energy difference between neighboring levels is small, rendering the electron energy spectrum nearly continuous and resembling that of free electrons. As the magnetic field strength increases, pronounced oscillations in $T_c({\bf B})$ emerge, as shown in the left inset of Figure 2.These oscillations continue at higher magnetic fields; however, unlike the low-field regime, their amplitude significantly increases, with values both exceeding and falling below the BCS theory prediction of 1 K. This phenomenon is attributed to the fact that as the magnetic field increases, the number of LLs around the Fermi surface decreases, while the density of states within each LL increases, leading to an enhanced Cooper pair formation. The behavior of $T_c({\bf B})$ becomes particularly intriguing at very high magnetic fields, where it not only exhibits oscillations but also persists within specific magnetic field intervals, indicating a re-entrant superconducting phase. At these high field strengths, $T_c({\bf B})$ demonstrates a complex and compelling pattern, characterized by both pronounced oscillations and sustained non-superconductivity within distinct magnetic field ranges, as depicted in the right inset of Figure 2. This strong-field superconducting re-entrant behavior observed here is similar to the findings from the original semiclassical approximation\cite{Rasolt-1991}.

Having established the influence of LLs alone, we continued to investigate the influence of Zeeman splitting on the superconducting transition temperature ($T_c({\bf B})$ ) in a magnetic field by varying the $g$-factor. For the purpose of simplifying the analysis, we assumed that the energy splitting between adjacent LLs with opposite spins is less than 2 (in units of $\hbar\omega$, where $\omega$ is the cyclotron frequency), which corresponds to a $g$-factor less than 4. Figures 3 and 4 present the $T_c({\bf B})$ results for various $g$-factors, assuming spin-singlet pairing and Cooper pairs formed on the same and neighboring LLs, respectively.
\begin{figure}
\centering
{\includegraphics[width=0.5\textwidth]{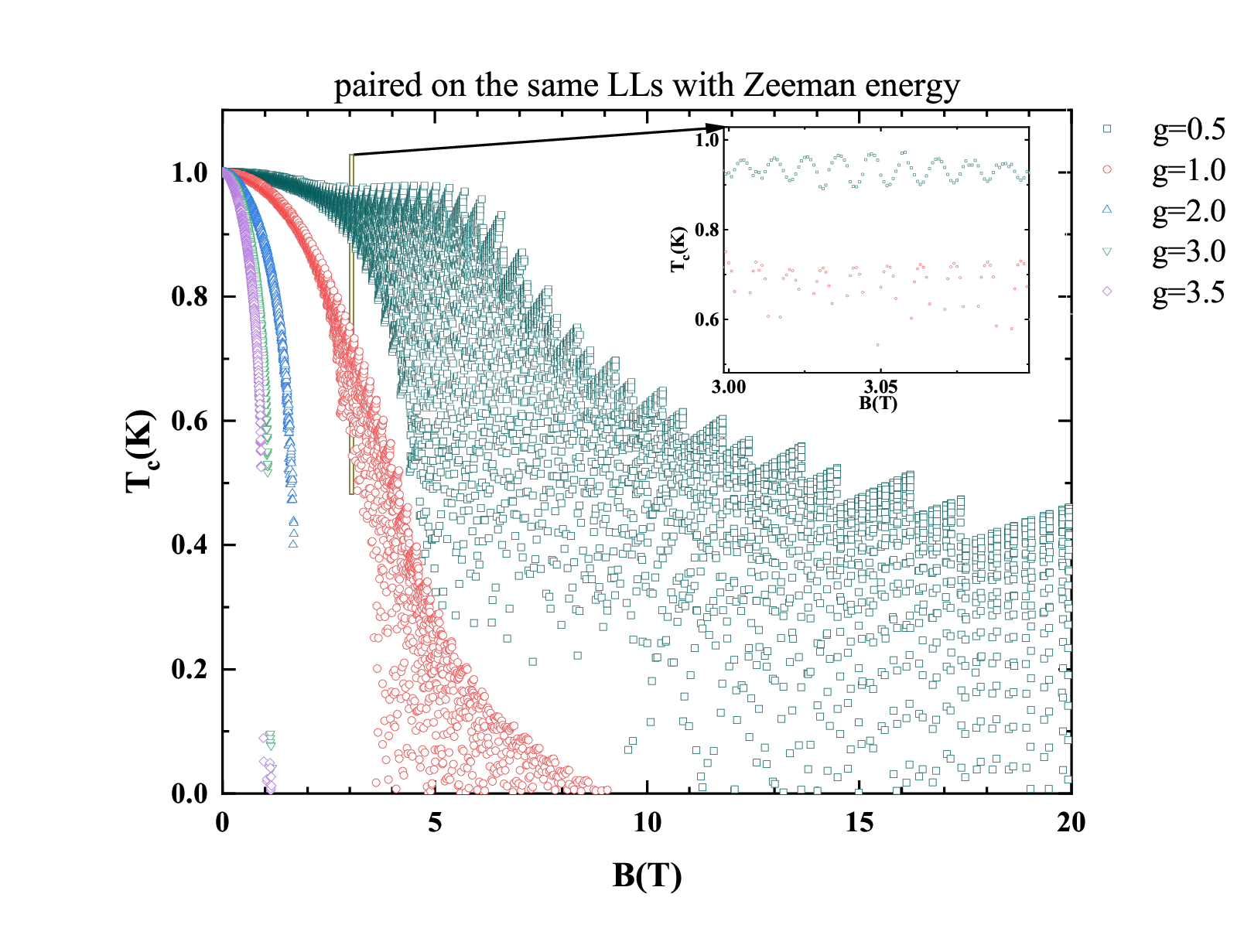}
\caption{The critical temperature, $T_c({\bf B})$, in a magnetic field, for electron pairing on the same LL with Zeeman splitting, where the electrons have differing $g$-factors. In this figure, $T_c({\bf B})$is represented by empty olive-green squares for $g = 0.5$, empty red circles for $g = 1.0$, empty blue triangles for $g = 2.0$, empty regular triangles for $g = 2.0$, empty inverted triangles for $g = 3.0$, and empty purple diamonds for $g = 3.5$. The inset clearly illustrates the Zeeman-splitting-induced oscillations of $T_c({\bf B})$ for small $g$-factors. }
\label{fig3}}
\end{figure}

Figure 3 illustrates the magnetic field dependence of the critical temperature, $T_c({\bf B})$, for electrons paired on the same LLs. At small $g$-factors, $T_c({\bf B})$ exhibits oscillatory behavior similar to that observed without the Zeeman splitting, with the oscillation amplitude increasing as the magnetic field strength increases. The inset of Figure 3 clearly shows these $T_c({\bf B})$ oscillations in magnetic fields. However, a key difference from the case without the Zeeman splitting (Figure 2) is that $T_c({\bf B})$ shows an overall decrease with increasing magnetic field. The behavior of $T_c({\bf B})$ becomes particularly intriguing at high magnetic fields, especially for small $g$-factors, where superconductivity re-emerges under strong magnetic fields, albeit with a significantly narrowed oscillation range of the re-emergent superconductivity, as shown in Figure 5a. This re-emergence mirrors the behavior of $T_c({\bf B})$ in the absence of the Zeeman effect. The data presented in Figure 3 demonstrate that the Zeeman energy (or $g$-factor) significantly influences the $T_c({\bf B})$ behavior in a magnetic field, with this influence being particularly pronounced at high magnetic field strengths.

\begin{figure}
\centering
{\includegraphics[width=0.5\textwidth]{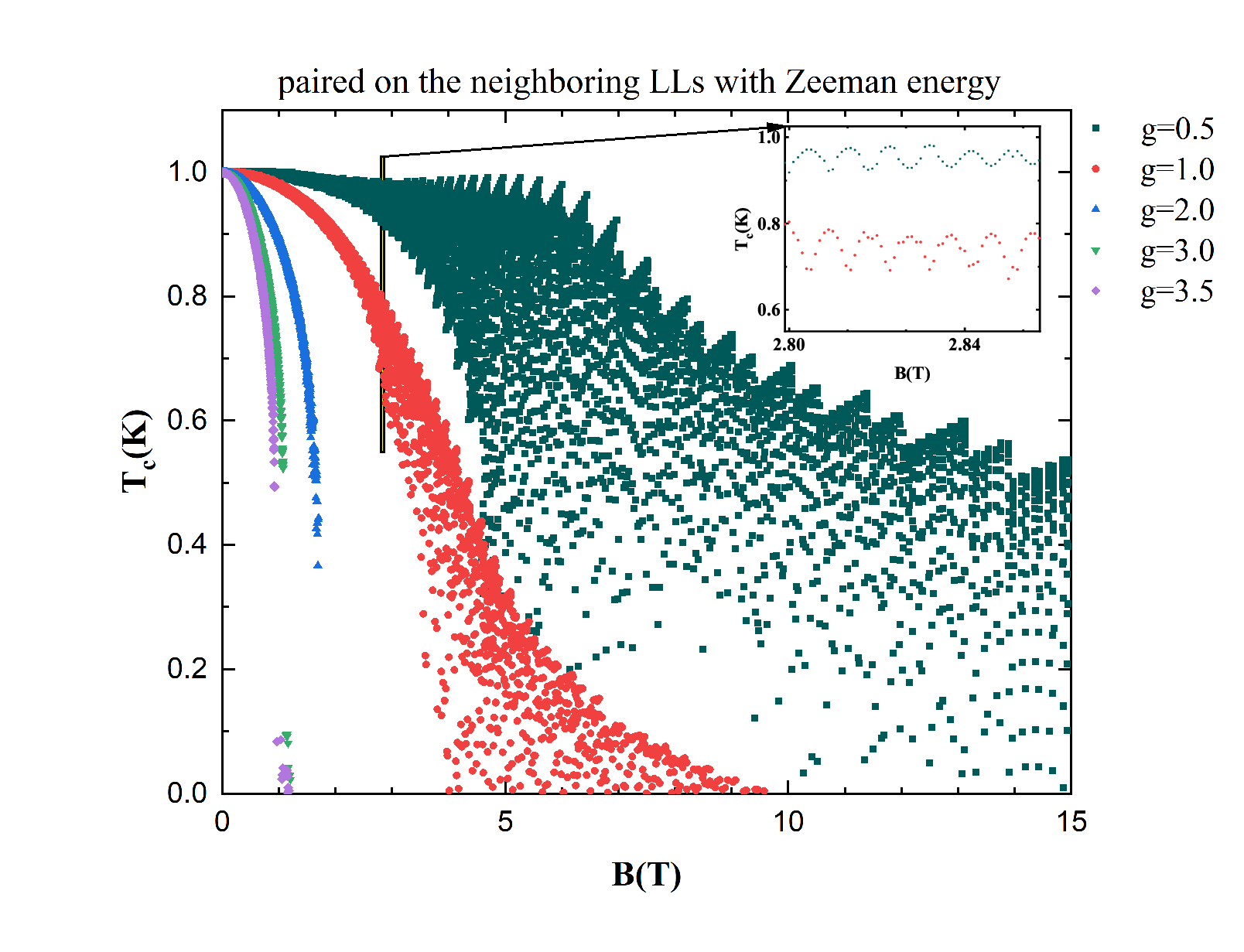}
\caption{The critical temperature, $T_c({\bf B})$, in a magnetic field, for electron pairing on the neighboring LLs with Zeeman splitting, where the electrons have differing $g$-factors. In this figure, $T_c({\bf B})$ is represented by full olive-green squares for $g = 0.5$, full red circles for $g = 1.0$, full regular triangles for $g = 2.0$, full inverted triangles for $g = 3.0$, and full purple diamonds for $g = 3.5$. The inset clearly illustrates the Zeeman-splitting-induced oscillations of $T_c({\bf B})$ for electron pairing on the neighboring LLs with small $g$-factors in a magnetic field. }
\label{fig4}}
\end{figure}

Figure 4 illustrates the critical temperature $T_c({\bf B})$ as a function of the magnetic field for electrons paired on the neighboring LLs. The data reveal a strong resemblance to the results for electrons paired on the same LLs. Additionally, this figure confirms that the Zeeman energy significantly affects the $T_c({\bf B})$ behavior in a magnetic field for electrons paired on the neighboring LLs. Furthermore, it is observed that in 2D superconductors, Cooper pairs formed on the same and neighboring LLs exhibit minimal influence on the superconducting critical temperature in a magnetic field, a contrast to the behavior seen in 3D systems \cite{Zhao-2025}.

\begin{figure}
\centering
{\includegraphics[width=0.45\textwidth]{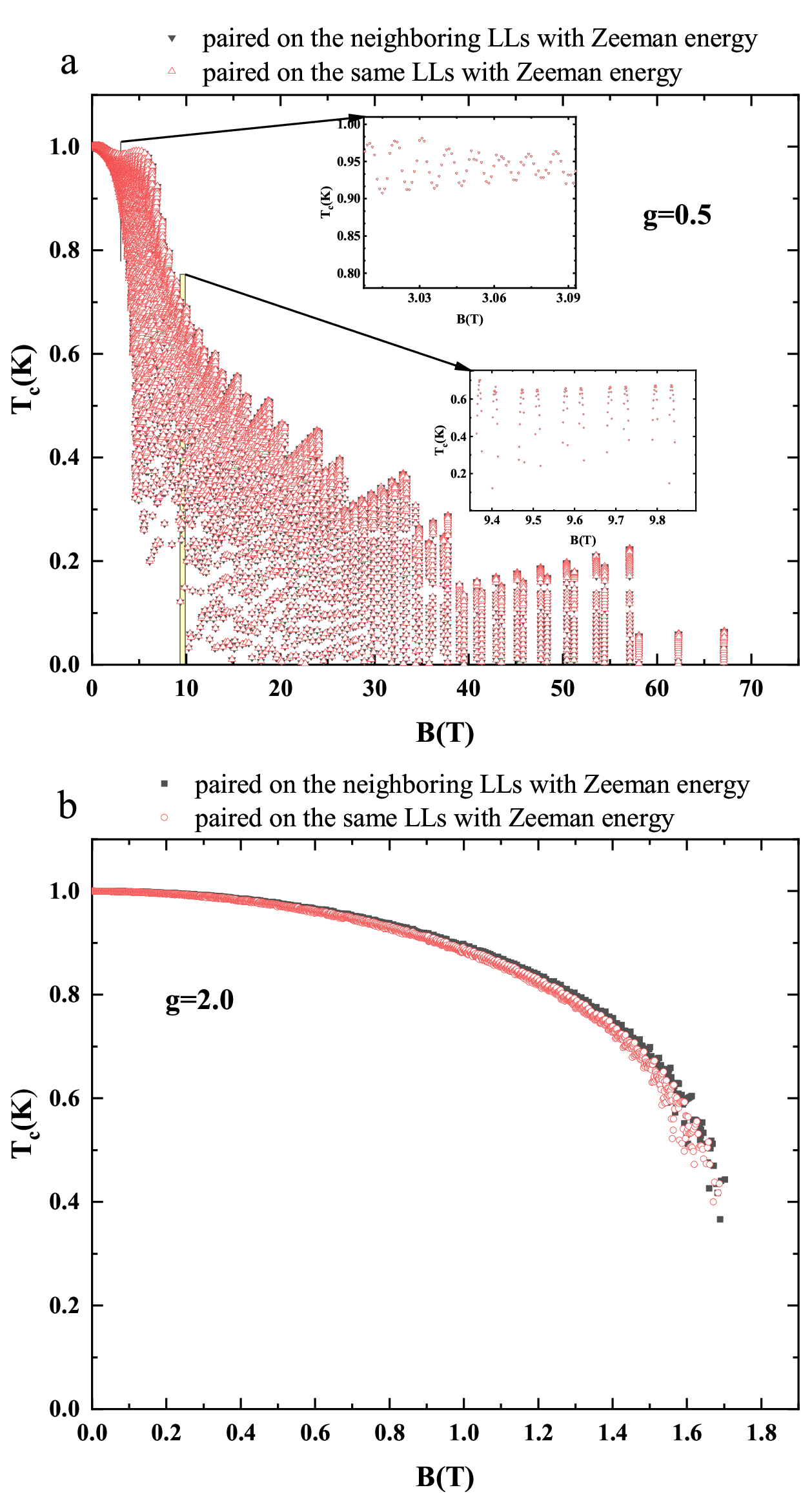}
\caption{
The magnetic field dependence of the critical temperature, $T_c({\bf B})$, reveals electron pairing in the same and neighboring LLs, considering Zeeman splitting with $g$-factors of (a) $g=0.5$ and (b) $g=2.0$. Figure (a) illustrates $T_c({\bf B})$, for electron pairing within the same (filled black inverted triangles) and neighboring (empty red triangles) LLs, with Zeeman splitting and $g=0.5$ applied. The inset in Figure (a) clearly demonstrates the Zeeman-splitting-induced oscillations of $T_c({\bf B})$ at both low and high magnetic fields. Figure (b) depicts $T_c({\bf B})$ for electron pairing on the same (filled black squares) and neighboring (empty red circles) LLs, with Zeeman splitting and $g=2.0$ applied.
 }
\label{fig5}}
\end{figure}

Figure 5 elucidates the distinct magnetic field dependence of the critical temperature, $T_c({\bf B})$, for electron paired on the same and neighboring LLs, accounting for Zeeman splitting with $g$-factors of $g=0.5$ and $g=2.0$. Specifically, in Figure 5a, with $g=0.5,$ $T_c(B)$ exhibits oscillatory behavior across the magnetic field range, and notably, a re-entrant superconducting phase emerges at very high magnetic fields. The inset in Figure 5a clearly demonstrates these Zeeman-splitting-induced oscillations of $T_c({\bf B})$ at both low and high magnetic fields. While the oscillation amplitude varies with the magnetic field, a clear pattern is not discernible. In contrast, as shown in Figure 5b, representing $g=2.0$, the re-entrant superconducting phase at high magnetic fields and the disappearance of $T_c({\bf B})$ oscillations are observed.

Figures 3-5 illustrate that the superconducting transition temperature, $T_c({\bf B})$, exhibits a marked decrease with increasing $g$-factor in a magnetic field, underscoring the significant contribution of the Zeeman effect. This result aligns with experimentally determined critical magnetic field behavior, emphasizing the necessity to incorporate the Zeeman term's effects in magnetic field analyses. Furthermore, as shown in Figure 5, a negligible difference in $T_c({\bf B})$ is observed between Cooper pairs formed on the same and neighboring LLs, attributed to the high density of LLs near the FS. This is further compounded by the absence of third-dimensional constraints in 2D superconductors, unlike their 3D counterparts, resulting in a less diverse $T_c({\bf B})$ behavior in 2D systems.

This study is limited to the clean limit. Nevertheless, impurities can substantially alter superconducting properties, including the upper critical field and LL broadening, thereby affecting $T_c({\bf B})$. Future research will investigate impurity effects on mixed electron pairing on the same  and neighboring LLs, focusing on the critical temperature, upper critical field, and the gap in Type-I superconductors. This work should also apply to 2D Type-II superconductors below  the lower critical field $H_{c1}$ \cite{KC-1993}, so that there are no vortices.  $H_ {c1}$  first needs to be calculated for Type-II superconductors in 2D with the  pairing of electrons in a magnetic field on the same and on different LLs.  This could be done using the GL model approach, as suggested for 3D superconductors previously \cite{Zhao-2025} Eventually, yet future work could examine vortex behavior using these or more sophisticated theoretical frameworks.

\section{Conclusions}
In the present paper, we analyzed the electronic behavior within momentum space and focused on the critical point where the system between the normal and superconducting states in the presence of a magnetic field, by following the BCS theory approach to investigate the effects of LLs and Zeeman splitting on the 2D superconducting critical temperature. These are different from those obtained by treating the effects of the magnetic field semiclassically using the phase factor $\exp({\frac{ie}{\hbar}\int_{\bm{r}_1}^{\bm{r}_2}\bm{A}\cdot d\bm{r}})$. We expressed the energy of the paired electrons in an external magnetic field as the quantized energy levels $(n+\frac{1}{2})\hbar\omega+\frac{g}{2}\mu_{B} \bm{\sigma}\cdot\bm{B}-\mu_F$, which have a high level of degeneracy for electrons paired in orbits perpendicular to the magnetic field. The quantized LLs, when influenced by the Zeeman energy $\frac{g}{2}\mu_{B} \bm{\sigma}\cdot\bm{B}$, allow the possibility of the electrons forming Cooper pairs on the same or neighboring LLs with opposite spins in the direction of the magnetic field.

We found that, at low magnetic fields and without spin response, Cooper pairs formed on the same LLs induce an oscillatory $T_c({\bf B})$ around the BCS prediction. With increasing magnetic field, $T_c({\bf B})$ continues to oscillate, crossing the BCS prediction. At very high magnetic fields, $T_c({\bf B})$ exhibits oscillations but persistent superconducting behavior within specific field intervals. Furthermore, $T_c({\bf B})$ displays minimal variation between electron pairing on the same and neighboring LLs in magnetic fields with Zeeman splitting. Notably, $T_c({\bf B})$ gradually decreases, and oscillations diminish with increasing $g$-factor for pairing on the same and neighboring LLs. Most significantly, a re-entrant superconducting phase emerges at very high magnetic fields with small $g$-factors, resembling pairing within the same LLs without Zeeman splitting.

\begin{acknowledgments}
 This work was supported by the National Natural Science Foundation of China through Grant No. 11874083. 
\end{acknowledgments}

\end{document}